\documentclass{eds2019}

\def\Journal#1#2#3#4{{#1} {\bf #2}, #3 (#4)}

\def\NIMA{{\em Nucl. Instrum. Methods} A}
\def\NPB{{\em Nucl. Phys.} B}
\def\PLB{{\em Phys. Lett.}  B}
\def\PRL{\em Phys. Rev. Lett.}
\def\PRD{{\em Phys. Rev.} D}

\def\JHEP{\em J. High Energy Phys.}

\def\be{\begin{equation}}
\def\ee{\end{equation}}
\def\bea{\begin{eqnarray}}
\def\eea{\end{eqnarray}}

\newcommand{\Photo}{}

\begin{document}
\vspace*{4cm}
\title{Gluon polarization measurements from longitudinally polarized proton-proton collisions at STAR}

\author{Zilong Chang for the STAR Collaboration}

\address{Department of Physics, Brookhaven National Laboratory,\\
Upton, NY 11973}

\maketitle\abstracts{
Jets produced in the pseudo-rapidity range, $-1.0 < \eta < 1.0$, from $pp$ collisions at RHIC kinematics are dominated by quark-gluon and gluon-gluon scattering processes. Therefore the longitudinal double spin asymmetry $A_{LL}$ for jets is an effective channel to explore the longitudinal gluon polarization in the proton. At STAR, jets are reconstructed in full azimuth, from the charged-particle tracks seen by the Time Projection Chamber and electro-magnetic energy deposited in the Barrel and Endcap electro-magnetic calorimeters at both $\sqrt{s} = $ 200 and 510 GeV. Early STAR inclusive jet $A_{LL}$ results at $\sqrt{s} = $ 200 GeV provided the first evidence of the non-zero gluon polarization at $x > $ 0.05. At  $\sqrt{s} = $ 510 GeV, the inclusive jet $A_{LL}$ is sensitive to the gluon polarization as low as $x \sim $ 0.015. In this talk, we will discuss recent STAR inclusive jet and dijet $A_{LL}$ results at $\sqrt{s} = $ 510 GeV and highlight the new techniques designed for this analysis, for example the underlying event correction to the jet transverse energy and its effect on the jet $A_{LL}$. Dijet $A_{LL}$ results are shown for four topologies in regions of pseudo-rapidity, effectively scanning the $x$-dependence of the gluon polarization.     
}

\section{Introduction}\label{sec:intro}

The spin of the proton has been known as $\frac{1}{2}$, however the early deep inelastic scattering, DIS, experiments showed that quarks inside the proton contribute about 30\% of its total spin \cite{emc1988}. Where the rest of the proton spin comes from has been puzzling for the past three decades. Neither inclusive or semi-inclusive measurements from DIS experiments offer a good description of the gluon contribution \cite{lss10}.

\begin{figure}[h]
  \centering
  \includegraphics[width=0.5\columnwidth]{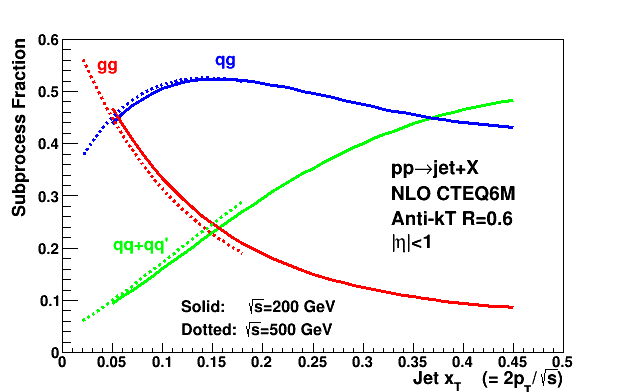}
  \caption[]{Fraction of jet production due to $qq$, $qg$ and $gg$ processes as a function of jet $x_T = \frac{2p_T}{\sqrt{s}}$ from the next-to-leading order, NLO, perturbative QCD calculations at $\sqrt{s} =$ 200 and 500 GeV \cite{nlo2012}.}
  \label{fig:subp}
\end{figure}

Different from DIS experiments, hadronic collisions such as proton-proton, $pp$, collisions provide an effective environment to study the gluon polarization inside the proton. In longitudinally polarized $pp$ collisions, the double spin asymmetry, $A_{LL}$, for jets, defined as the fractional difference of the jet cross section when two beams have the same and the opposite helicities, is proportional to the partonic double spin asymmetry, $\hat{a}_{LL}$, convoluted with the polarized parton distribution functions, PDFs, of the scattering partons. At center of mass energy $\sqrt{s} =$ 200 and 510 GeV, the jet production is contributed by hard quantum chromodynamics, QCD, scattering processes. As seen from Figure \ref{fig:subp}, $qg$ and $gg$ processes dominate the jet productions at jet $x_T = \frac{2p_T}{\sqrt{s}} < \sim 0.3$, where the cross section is appreciable \cite{nlo2012}. Both processes have sizable partonic $\hat{a}_{LL}$, therefore jet $A_{LL}$ is sensitive to gluon polarization.

The Relativistic Heavy Ion Collider, RHIC, is the one and only hadronic collider capable of colliding polarized proton beams up to $\sqrt{s} =$ 510 GeV \cite{rhic,rhicdesign,rhicpol}. A hydrogen gas-jet polarimeter \cite{hjet} and a Coulomb nuclear interference proton-Carbon polarimeter \cite{pCpol} located at 12 o'clock in the ring provide the beam polarizations. The Solenoidal Tracker at RHIC, STAR, sitting at 6 o'clock is a symmetric detector in full azimuth, $0 < \phi < 2\pi$ \cite{star}, including, a time projection chamber, TPC, at $|\eta| < $ 1.3 \cite{tpc}, barrel and endcap electromagnetic calorimeters, BEMC and EEMC, $|\eta| < 1.0$ and $1.0 < \eta < 2.0$ respectively, \cite{bemc,eemc} and a forward meson spectrometer, FMS, at $2.6 < \eta < 3.9$. In addition, the vertex position detector, VPD \cite{vpd}, and the zero degree calorimeter, ZDC \cite{zdc}, are used to provide relative luminosity and local polarimetry monitoring.

STAR has published a series of results on both inclusive and dijet $A_{LL}$ over the past two decades at $\sqrt{s} = 200$ GeV \cite{run9aLL,run9bdj,run9edj}. The DSSV group who included the two recent STAR dijet $A_{LL}$ results, one with both jets in $|\eta| < 0.8 $ \cite{run9bdj} and the other with at least one jet in $0.8 < \eta < 1.8$ \cite{run9edj}, shows the gluon polarization, $\int_{0.01}^{1}\Delta g(x) = 0.296 \pm 0.108$ at $Q^2 =$ 10 $\textrm{GeV}^2$ \cite{dssv2019}.

\section{Jet reconstruction from 510 GeV data} \label{sec:jet510}

In the years 2012 and 2013, STAR has taken data from 510 GeV longitudinally polarized $pp$ collisions. The beam polarizations for both years are between 50\% to 53\% \cite{polnote}. The integrated luminosity for the 2013 run is about 3.7 times larger than that for the 2012 run, however the jet trigger in the 2013 run is optimized for the dijet measurements.

Jets are reconstructed from charged particles measured by the TPC and electro-magnetic energy collected by BEMC and EEMC towers, a $0.05 \times 0.05$ space in $\eta-\phi$. Due to large pile-up event and soft background environment at $\sqrt{s} = 510$ GeV, anti-$k_T$ algorithm with jet parameter $R =$ 0.5 was used in this analysis \cite{antikt}.

A technique called the off-axis cone method was introduced to correct the jet transverse energy due to underlying event activities \cite{cones}. The method collects the particles inside two off-axis cones centered at the jet $\eta$ and $\pm \frac{\pi}{2}$ off the jet $\phi$ and finds the average transverse energy density of the two cones. The correction, $dp_T$, is taken as the product of the jet area and the found cone energy density. It allows one to study the $\eta$ dependence of the underlying event activities. Furthermore with this method, the underlying event contribution to the jet $A_{LL}$ can be estimated.

In order to study the systematic uncertainty, an embedding sample was generated by embedding simulated QCD events into zero-bias events, randomly taken during the real collisions. GEANT was used to simulate the detector response \cite{geant}. Intermediate particles and partons in the simulation are saved as well to quantify the correction from the measured detector jet quantity to the parton jet quantity, which facilitates comparisons with theoretical calculations.

\begin{figure}[h]
  \centering
  \includegraphics[width=0.4\columnwidth]{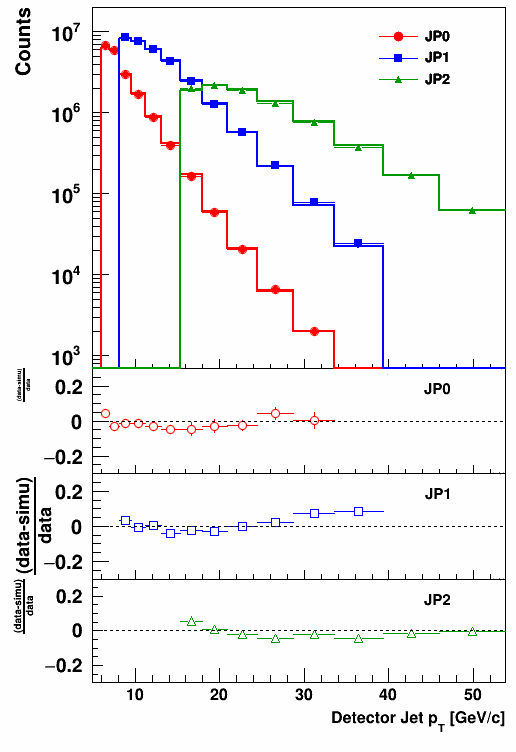}
  \caption[]{Comparison of jet $p_T$ spectrum for jets satisfying three jet patch triggers separately between data (blue points) and embedding (red lines) \cite{run12aLL}.}
  \label{fig:jetpt}
\end{figure}

The Perugia 2012 tune from PYTHIA6 \cite{pythia,perugia} overestimates the charged pion, $\pi^{\pm}$, yields at low transverse momentum, $p_T$, comparing to previously published STAR measurements \cite{pipm2006,pipm2012}. By reducing the PARP(90) parameter from 0.24 by default to 0.213, which controls the $\sqrt{s}$ dependence of the cut-off $p_{T,0}$, where it regulates the low $p_T$ divergence of the partonic cross section, a better agreement is achieved \cite{adkins,chang}. Figure \ref{fig:jetpt} shows an excellent agreement between data and embedding for the jet transverse momentum, $p_T$ spectrum satisfying three jet-patch triggers separately \cite{run12aLL}.

\section{Inclusive jet and dijet $A_{LL}$}\label{sec:aLL}
Experimentally, the jet $A_{LL}$ is calculated from the difference in the number of jets measured when beams have the same and the opposite helicities, $N^{++}$ and $N^{+-}$ respectively, by taking into account the relative luminosity, $R$ and beam polarizations, $P_1$ and $P_2$, as in Eq. \ref{eq:aLL}:

\be
A_{LL} = \frac{1}{P_1P_2}\frac{N^{++} - RN^{+-}}{N^{++} - RN^{+-}}.
\label{eq:aLL}
\ee

\begin{figure}[h]
  \centering
  \includegraphics[width=0.5\columnwidth]{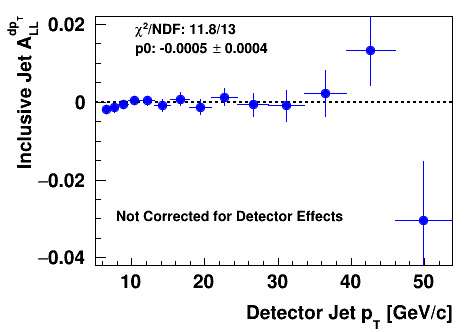}
  \caption[]{Underlying event $A_{LL}^{dp_T}$ as a function of detector jet $p_T$with a constant fit \cite{run12aLL}.}
  \label{fig:aLLdpt}
\end{figure}

Similarly the longitudinal double spin asymmetry for the underlying event correction $dp_T$, $A_{LL}^{dp_T}$, can be defined as the fractional difference of the averaged $dp_T$ for jets found when beams have the same and the opposite helicities as in Eq. \ref{eq:aLLdpt}. Figure \ref{fig:aLLdpt} shows the $A_{LL}^{dp_T}$ is consistent with zero \cite{run12aLL}. Its numerical value is used to calculate the potential impact the underlying events have on the longitudinal double spin dependent jet cross section, $\sigma^{++}$ and $\sigma^{+-}$. It turns out the the underlying event activity contribution to the jet $A_{LL}$ is at the level of $10^{-4}$, assigned as a systematic uncertainty.

\be
A_{LL}^{dp_T} = \frac{1}{P_1P_2} \frac{<dp_T>^{++} - <dp_T>^{+-}}{<dp_T>^{++} + <dp_T>^{+-}}
\label{eq:aLLdpt}
\ee

\begin{figure}[h]
  \begin{minipage}{0.5\linewidth}
    \centerline{\includegraphics[width=0.7\linewidth]{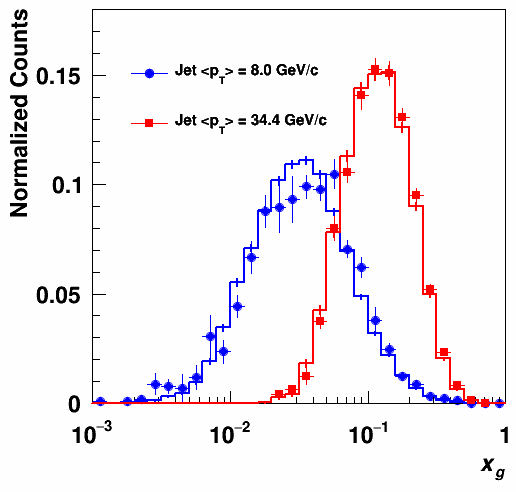}}
  \end{minipage}
  \hfill
  \begin{minipage}{0.5\linewidth}
    \centerline{\includegraphics[width=\linewidth]{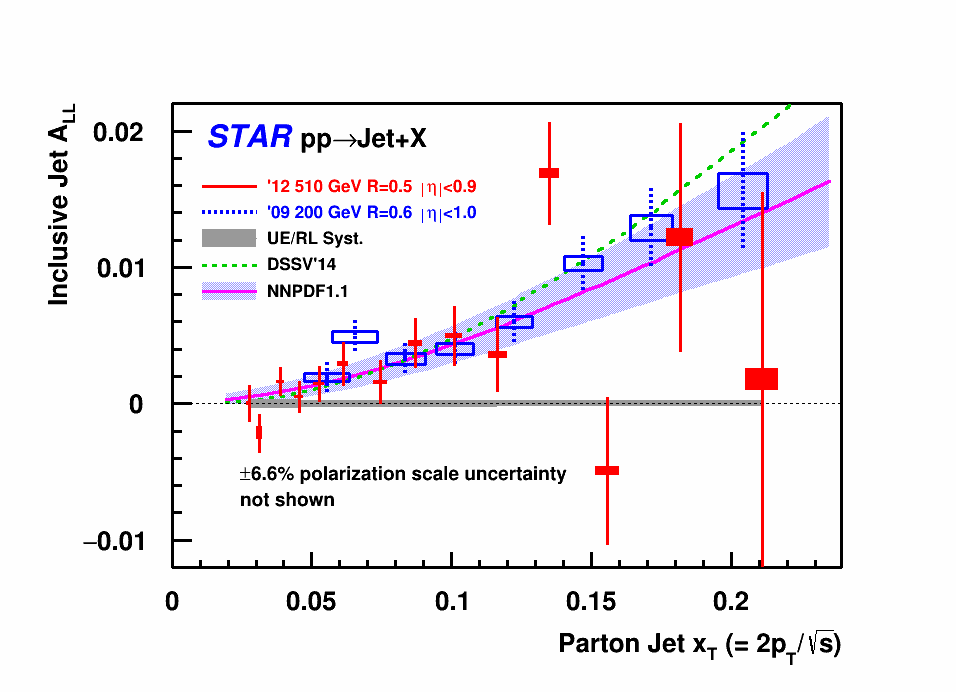}}
  \end{minipage}
  \caption[]{Left panel is the sampled $x_g$ distribution for jets in two $p_T$ bins. Right panel is the inclusive jet $A_{LL}$ vs. parton jet $x_T$ from STAR 2012 510 GeV data (red) and 2009 200 GeV data (blue) together with NLO polarized PDF predictions from DSSV (dashed line) \cite{dssv2014} and NNPDFpol1.1 \cite{nnpdf1.1} models (solid line with shades) \cite{run12aLL}.}
  \label{fig:aLL}
\end{figure}

The right panel of Figure \ref{fig:aLL} shows the inclusive jet $A_{LL}$ compared to the STAR 2009 results at $\sqrt{s} =$ 200 GeV \cite{run9aLL} as a function of jet $x_T$. The two results at different $\sqrt{s}$ agree well with each other, while the new techniques developed in this analysis lead to much reduced systematic uncertainty than the previous measurements. The 510 GeV results are also consistent with recent polarized PDF predictions \cite{dssv2014,nnpdf1.1}. As seen in the left panel of Figure \ref{fig:aLL} the sampled $x_g$ can reach as low as 0.015 \cite{run12aLL}.

\begin{table}[ht]
\caption[]{Definition of four topologies, forward-forward, forward-central, central-central, and forward-backward for dijets \cite{run12aLL}}
\label{tab:top}
\begin{center}
\begin{tabular}{|c|c|c|}
  \hline
  A & Forward-Forward& 0.3 $ < |\eta_{3,4}| < $ 0.9, $\eta_3 \cdot \eta_4 >$ 0\\
  \hline
  B & Forward-Central& $|\eta_{3,4}| < $ 0.3, 0.3 $ < |\eta_{4,3}| < $ 0.9\\
  \hline
  C & Central-Central& $|\eta_{3,4}| < $ 0.3 \\
  \hline
  D & Forward-Backward& 0.3 $ < |\eta_{3,4}| < $ 0.9, $\eta_3 \cdot \eta_4 <$ 0\\
\hline  
\end{tabular}
\end{center}
\end{table}

\begin{figure}[h]
  \begin{minipage}{0.5\linewidth}
    \centerline{\includegraphics[width=0.48\linewidth]{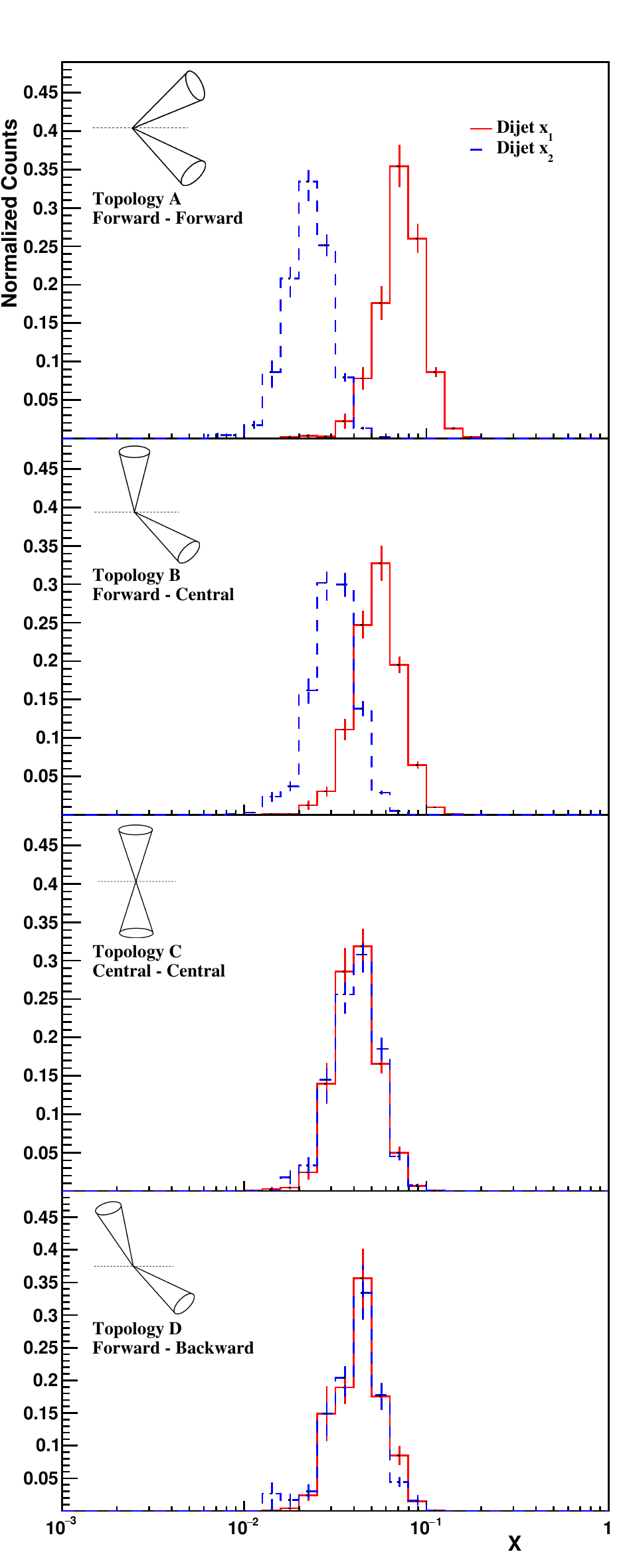}}
  \end{minipage}
  \hfill
  \begin{minipage}{0.5\linewidth}
    \centerline{\includegraphics[width=0.56\linewidth]{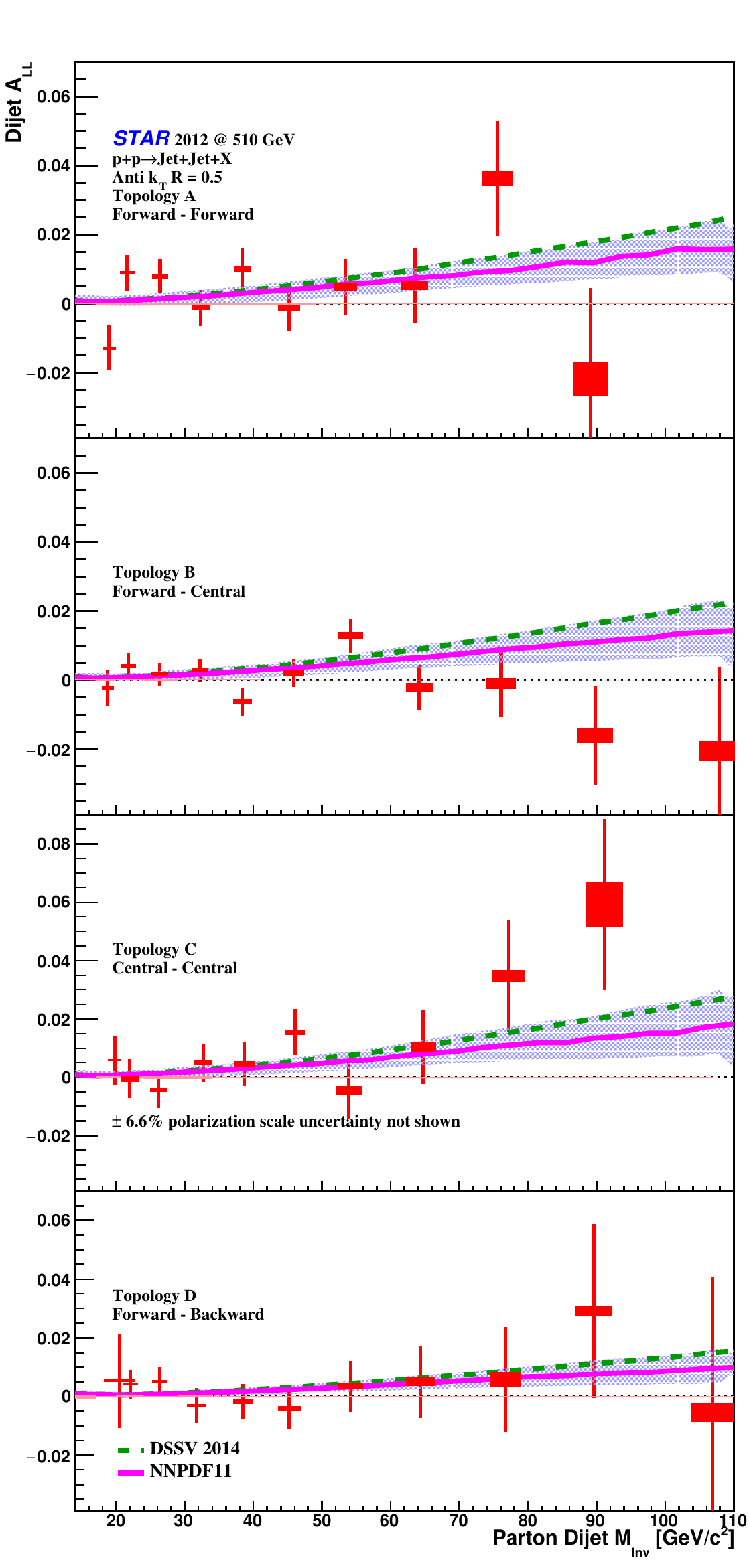}}
  \end{minipage}
  \caption[]{Left panels show the sampled $x_1$  and $x_2$ distributions by dijets for four topologies with invariant mass, $M_{inv} =$ 17 $-$ 20 $\textrm{GeV}/c^2$. Right panels show the dijet $A_{LL}$ vs. parton dijet $M_{inv}$ for four topologies from STAR 2012 510 GeV data (red) together with NLO polarized PDF predictions from DSSV (dashed line) \cite{dssv2014} and NNPDFpol1.1 \cite{nnpdf1.1} models (solid line with shades) \cite{run12aLL}.}
  \label{fig:aLLdj}
\end{figure}

The initial parton kinematics, $x_1$ and $x_2$, can be calculated from the kinematics of the measured dijet events. In this analysis, the dijet $A_{LL}$ are measured in four distinct topologies based on the $\eta$ range of the two jets, as listed in Table \ref{tab:top}. Each topology features its own sampled $x_1$ and $x_2$ distributions, therefore the sampled parton $x$ distributions are much narrower than that sampled by inclusive jets. The $A_{LL}$ also varies with topologies because the partonic $\hat{a}_{LL}$ depends on the scattering angle $cos\theta^*$. The dijet $A_{LL}$ and its sampled $x_1$ and $x_2$ distributions can be seen in Figure \ref{fig:aLLdj} \cite{run12aLL}.

The preliminary results of STAR 2013 inclusive and dijet $A_{LL}$ at $\sqrt{s} = 510$ GeV \cite{run13aLL} are consistent with the 2012 results. The systematic uncertainties are under further study for future publication.

\section{Other measurements and STAR forward upgrade} \label{sec:others}

$\pi^{0}$s are recontructed from FMS in two $\eta$ ranges, 2.65 $< \eta < 3.15$ and 3.15 $< \eta <$ 3.90. The inclusive $\pi^{0}$ $A_{LL}$ as a function of transverse momentum is found to be very small, less than $5 \times 10^{-3}$. The $A_{LL}$ will help to constrain the gluon polarization at $x_g \sim 10^{-3}$ \cite{fmspi0}.

STAR is preparing for an upgrade in the forward region, 2.8 $< \eta <$ 3.7, which requires the installation of a forward calorimeter and a forward tracking system in time for the 2022 RHIC run \cite{fwdupg}. The calorimeter, placed at the current FMS location, includes both a hadron calorimeter and an electro-magnetic calorimeter. Three silicon ministrip detectors combined with four small strip thin gap chambers, sTGC, provide the tracking system. The dijet $A_{LL}$ will be one of the highlighted measurements for the upgrade with one or both of the two jets in the forward region. The predicted precision will be around 5 to 6 times better than the current uncertainty band at invariant dijet mass less than 20 GeV/$c^2$ when both jets are detected in the forward region.
\section{Conclusion} \label{sec:concl}
STAR inclusive jet and dijet $A_{LL}$ measurements are unique and can be used to unravel the polarized gluon PDF in the proton. The inclusive jet measurements probe the magnitude of the gluon polarization, while the dijet measurements constrain the shape of $\Delta g(x)$. The previous results at $\sqrt{s} =$ 200 GeV provided the first evidence of positive gluon polarization at $x >$ 0.05, furthermore at $\sqrt{s} =$ 510 GeV, the jet $A_{LL}$ will enable access to $\Delta g(x)$ at $x$ around 0.015. We expect future results with more precision at both $\sqrt{s}$. In the near future, the forward upgrade will explore small $x$ gluon polarization as low as $10^{-3}$.

\end{document}